\documentclass{desyprocA4}

\usepackage{graphicx}
\usepackage{amsmath}
\usepackage{latexsym}
\usepackage{slashbox}
\usepackage{array}
\usepackage{dcolumn}
\usepackage{bm}
\usepackage{epsfig}
\usepackage{cite}
\usepackage{multirow}

\usepackage[dvips]{color}
\definecolor{Black}{named}{Black}
\definecolor{Red}{named}{Red}
\definecolor{Blue}{named}{Blue}

\def\lsim{\mathrel{\rlap{\raise 2.5pt \hbox{$<$}}\lower 2.5pt
\hbox{$\sim$}}}

\newcommand{\dd}{\mbox{{\rm d}}}

\newcommand{\Lumint}{{\cal L}_{\rm int}}

\def\epem{\ifmmode e^+e^-\else $e^+e^-$\fi}
\def\to{\rightarrow}

\def\mpl{\ifmmode \overline M_{Pl}\else $\bar M_{Pl}$\fi}
\def\beq{\begin{equation}}
\def\be{\begin{equation}}
\def\beqn{\begin{eqnarray}}
\def\ee{\end{equation}}
\def\eeq{\end{equation}}
\def\eeqn{\end{eqnarray}}

\begin{document}


\title{Sneutrino Identification in Lepton Pair Production at ILC with Polarized Beams}

\author{{\slshape A.~V. Tsytrinov$^1$}, J. Kalinowski$^{2}$, P. Osland$^{3}$, A.~A. Pankov$^1$\\[1ex]
$^1$The Abdus Salam ICTP Affiliated Centre, Technical University
of Gomel, 246746 Gomel, Belarus\\
$^2$Faculty of Physics, University of Warsaw,
  Ho{\.z}a 69, 00-681 Warsaw, Poland\\
$^3$Department of Physics and Technology, University of Bergen, Postboks 7803, N-5020  Bergen, Norway}


\contribID{xy}

\confID{1964}  
\desyproc{DESY-PROC-2012-123}
\acronym{PLHC2010} 
\doi  

\maketitle

\date{\today}

\begin{abstract}
  Numerous non-standard dynamics are described by
contact-like effective interactions that can manifest themselves
in electron-positron collisions
only through deviations of the observables (cross sections,
asymmetries) from the Standard Model predictions. If such
a deviation were observed, it would be important to
identify the actual source among the possible non-standard
interactions as many different new physics scenarios may lead to
very similar experimental signatures. We study the
possibility of uniquely identifying the indirect
effects of  $s$-channel sneutrino  exchange, as predicted by supersymmetric theories
with $R$-parity violation, against other new physics scenarios in
high-energy $e^+e^-$ annihilation into lepton pairs at the
International Linear Collider. These competitive models are
interactions based on gravity in large and in TeV-scale
extra dimensions, anomalous gauge couplings, $Z'$ vector bosons
and compositeness-inspired four-fermion contact interactions.
To evaluate the identification reach on sneutrino exchange,
we use as basic observable a double polarization
asymmetry, that is particularly suitable to directly test for such
$s$-channel sneutrino exchange effects in the data analysis. The
availability of both beams being polarized plays a crucial r{\^o}le in
identifying the new physics scenario.
\end{abstract}


\section{Introduction}\label{sec:I}
Numerous new physics (NP) scenarios, candidates as solutions of
Standard Model (SM) conceptual problems, are characterized by
novel interactions mediated by exchanges of very heavy states with
mass scales significantly greater than the electroweak scale. In
many cases, theoretical considerations as well as current
experimental constraints indicate that the new objects may be too
heavy to be directly produced even at the highest energies of the
CERN Large Hadron Collider (LHC) and at foreseen future colliders,
such as the $e^+e^-$ International Linear Collider (ILC). In this
situation the new, non-standard, interactions would only be
revealed by indirect, virtual, effects manifesting themselves as
deviations from the predictions of the SM. In the case of indirect
discovery the effects may be subtle since many different NP
scenarios may lead to very similar experimental
signatures and they may easily be confused in certain regions of
the parameter space for each class of models.

At the available energies provided by the accelerators,
where we study reactions among the familiar SM particles,
effective contact interaction Lagrangians represent a
convenient theoretical tool to physically parameterize the effects
of the above-mentioned non-standard interactions and, in
particular, to test the corresponding virtual high-mass exchanges.
There are many very different NP scenarios that predict new
particle exchanges which can lead to contact interactions (CI) which may show up below
direct production thresholds. These are compositeness
\cite{Eichten:1983hw}, a $Z'$ boson from models with an extended gauge
sector
\cite{Langacker:2008yv,Rizzo:2006nw,Leike:1998wr,Hewett:1988xc},
scalar or vector leptoquarks \cite{Buchmuller:1986zs}, $R$-parity
violating sneutrino ($\tilde \nu$) exchange
\cite{Kalinowski:1997bc,Rizzo:1998vf}, bi-lepton boson exchanges
\cite{Cuypers:1996ia}, anomalous gauge boson couplings (AGC)
\cite{Gounaris:1997ft}, virtual Kaluza--Klein (KK) graviton
exchange in the context of gravity propagating in large extra
dimensions, exchange of KK gauge boson towers or string
excitations
\cite{ArkaniHamed:1998rs,Giudice:1998ck,Hewett:1998sn,Antoniadis:1993jp,Pasztor:2001hc,Cheung:2001mq},
{\it etc.}
Of course, this list is not exhaustive, because other kinds of
contact interactions may be at play.

If $R$-parity is violated it is possible that the exchange of
sparticles can contribute significantly to SM processes and may
even produce  peaks or bumps~{\cite
{Kalinowski:1997bc,Rizzo:1998vf}} in cross sections if they are
kinematically accessible. Below threshold, these new spin-0
exchanges may make their manifestation known via indirect effects
on observables (cross sections and asymmetries),
including spectacular decays \cite{Bomark:2011ye}.
Here we will
address the question of whether the effects of the exchange of
scalar (spin-0) sparticles can be differentiated  at linear colliders from those
associated with the wide class of other contact interactions
mentioned above.

For a sneutrino in an $R$-parity-violating theory, we take the
basic couplings to leptons and quarks to be given by
\begin{equation}
\lambda_{ijk} L_{i}L_{j}{\bar{E}}_{k}
+\lambda_{ijk}^{\prime}L_{i}Q_{j}{\bar{D}_{k}}. \label{Rviol}
\end{equation}
Here, $L$ $(Q)$ are the left-handed lepton (quark) doublet
superfields, and ${\bar{E}}$ (${\bar{D}}$) are the corresponding
left-handed singlet fields. If just the $R$-parity violating
 $\lambda LL{\bar{E}}$ terms of the superpotential are present
it is clear that observables associated with leptonic
processes
\begin{equation}
e^++e^-\to \mu^++\mu^- \quad(\text{or } \tau^-+\tau^+),
\label{proc}
\end{equation}
will be affected by the exchange of $\tilde \nu$'s in the $t$- or
$s$-channels \cite{Kalinowski:1997bc,Rizzo:1998vf}. For instance,
in the case only one nonzero Yukawa coupling is present,
$\tilde\nu$'s may contribute to, e.g. $e^+e^-\to\mu^+\mu^-$ via
$t$-channel exchange. In particular, if $\lambda_{121}$,
$\lambda_{122}$, $\lambda_{132}$, or $\lambda_{231}$ are nonzero,
the $\mu^+\mu^-$ pair production proceeds via additional
$t$-channel sneutrino exchange mechanism. However, if only the
product of Yukawa, e.g.\ $\lambda_{131}\lambda_{232}$, is nonzero
the $s$-channel $\tilde\nu_\tau$ exchange would contribute to the
$\mu^+\mu^-$ pair final state. Below we denote by $\lambda$ the
relevant Yukawa coupling from the superpotential (\ref{Rviol})
omitting the subscripts.

In this note, we discuss the deviations induced by the
$s$-channel sneutrino exchange and contact interactions in
electron--positron annihilation into lepton pairs (\ref{proc})
at the planned ILC.  In particular, we use as a basic observable a
double polarization asymmetry that will unambiguously
identify $s$-channel sneutrino exchange effects
 in the data, relying on its spin-0 character and by
\emph{filtering} out contributions of other NP interactions.\footnote{This approach was earlier exploited for the discrimination against $Z^\prime$ exchange \cite{Rizzo:1998vf}.}
The availability of both beams being polarized plays a crucial r{\^o}le in
identifying that new physics scenario \cite{MoortgatPick:2005cw}.
 On the other hand, we note
that if only single (electron) beam polarization is available, the
left-right asymmetry does not help to unambiguously identify an $s$-channel
sneutrino exchange signature.\footnote{For the case of single beam polarization, $A_\text{LR}$ is an analogue of $A_\text{double}$ defined by Eq.~(\ref{double}).}

The  $R$-parity violating $s$-channel sneutrino exchange  in the process (\ref{proc}) requires a non-zero coupling $\lambda_{131}$ ($\lambda_{121}$).  This would necessarily induce non-standard contributions to Bhabha scattering,
\begin{equation}
e^++e^-\to e^++e^-,
\label{bhabha}
\end{equation}
which we also study, in order to compare the sensitivities in these channels.

We also compare the capability of the ILC to distinguish
effects of $s$-channel sneutrino exchange in the lepton pair
production process from other NP interactions with the corresponding
potential of the Drell-Yan process ($l=e,\mu$)
\cite{Osland:2010yg}
\begin{equation}
p+p\to l^++l^-+X  \label{proc_DY}
\end{equation}
at the LHC.

For completeness, we will in Sec.~\ref{sec:observ}
recall a minimum of relevant formulae defining the basic observables
used in our analysis.   In Sec.~\ref{sec:numeric} we perform the
numerical analysis, evaluating discovery and identification reaches
on sneutrinos. Finally, Sec.~\ref{sect:concl} contains some
concluding remarks.

\section{Observables and NP parametrization}\label{sec:observ}
We concentrate on the process
$e^++e^-\to \mu^+ +\mu^-$.
With $P^-$ and ${P^+}$ denoting the longitudinal polarizations of the electrons
and positrons, respectively, and $\theta$ the angle between the
incoming electron and the outgoing muon in the c.m.\ frame, the
differential cross section in the presence of  contact
interactions can be expressed as ($z\equiv\cos\theta$)
\cite{Schrempp:1987zy,Djouadi:1991sx}:
\begin{equation}
\frac{\dd\sigma^{\rm CI}}{\dd z} =\frac{3}{8} \left[(1+z)^2
{\sigma}_+^{\rm CI} +(1-z)^2 {\sigma}_-^{\rm CI}\right].
\label{cross}
\end{equation}
In terms of the helicity cross sections $\sigma_{\alpha\beta}^{\rm
CI }$ (with $\alpha,\beta={\rm L,R}$), directly related to the
individual CI couplings $\Delta_{\alpha\beta}$ (see Eq.~(\ref{amplit})):
\begin{eqnarray}
{\sigma}_{+}^{\rm CI}&=&\frac{1}{4}\,
\left[(1-P^-)(1+{P^+})\,\sigma_{\rm LL}^{\rm CI}
+(1+P^-)(1-{P^+})\,\sigma_{\rm RR}^{\rm CI}\right]\nonumber \\
&=&\frac{D}{4}\,\left[(1-P_{\rm eff})\,\sigma_{\rm LL}^{\rm CI}
+(1+P_{\rm eff})\,\sigma_{\rm RR}^{\rm CI}\right],
\label{s+} \\
{\sigma}_{-}^{\rm CI}&=&\frac{1}{4}\,
\left[(1-P^-)(1+{P^+})\,\sigma_{\rm LR}^{\rm CI}
+ (1+P^-)(1-{P^+})\,\sigma_{\rm RL}^{\rm CI}\right] \nonumber \\
&=& \frac{D}{4}\,\left[(1-P_{\rm eff})\,\sigma_{\rm LR}^{\rm CI}
+(1+P_{\rm eff})\,\sigma_{\rm RL}^{\rm CI}\right], \label{s-}
\end{eqnarray}
where the first (second) subscript refers to the chirality of the electron (muon) current.
Furthermore,
\begin{equation}
P_{\rm eff}=\frac{P^--{P^+}}{1-P^-{P^+}} \label{pg}
\end{equation}
is the effective polarization, $\vert P_{\rm eff}\vert\leq 1$, and
$D=1-P^- {P^+}$. For unpolarized positrons $P_{\rm eff}\rightarrow
P^-$ and $D\rightarrow 1$, but with ${P^+}\ne0$, $\vert P_{\rm
eff}\vert$ can be larger than $|P^-|$. Moreover, in
Eqs.~(\ref{s+}) and (\ref{s-}):
\begin{equation}
\sigma_{\alpha\beta}^{\rm CI}=\sigma_{\rm pt} \vert{\cal
M}_{\alpha\beta}^{\rm CI}\vert^2, \label{helcross}
\end{equation}
where  $\sigma_{\rm pt}\equiv\sigma(e^+e^-\to\gamma^\ast\to
\mu^+\mu^-) =(4\pi\alpha_\text{em}^2)/(3s)$. The helicity amplitudes ${\cal
M}_{\alpha\beta}^{\rm CI}$ can be written as
\begin{equation}
{\cal M}_{\alpha\beta}^{\rm CI}={\cal M}_{\alpha\beta}^{\rm
SM}+\Delta_{\alpha\beta} =
Q_eQ_\mu+g_\alpha^e\,g_\beta^\mu\,\chi_Z+ \Delta_{\alpha\beta},
\label{amplit}
\end{equation}
where
\begin{equation} \label{Eq:chi}
\chi_Z=\frac{s}{s-M^2_Z+iM_Z\Gamma_Z}
\end{equation}
represents the $Z$
propagator, $g_{\rm L}^l=(I_{3L}^l-Q_l s_W^2)/s_W c_W$ and $g_{\rm
R}^l=-Q_l s_W^2/s_W c_W$ are the SM left- and right-handed lepton
($l=e,\mu$) couplings of the $Z$ with $s_W^2=1-c_W^2\equiv
\sin^2\theta_W$ and $Q_l$ the leptonic electric charge. The
$\Delta_{\alpha\beta}$ functions represent the contact interaction
contributions coming from TeV-scale physics.

The structure of the differential cross section (\ref{cross}) is
particularly interesting in that it is equally valid for a wide
variety of NP models listed in Table~1.
Note that only graviton
and $t$-channel sneutrino exchanges induce a modified angular dependence to the differential
cross section via the  $z$-dependence of $\Delta_{\alpha\beta}$.

\begin{table}[tb]
\begin{center}
\begin{tabular}{|c|c|}
\hline Model & $\Delta_{\alpha\beta}$  \\  \hline\hline composite
fermions \cite{Eichten:1983hw} &
${\displaystyle{\pm\frac{s}{\alpha_\text{em}}\frac{1}
{\Lambda_{\alpha\beta}^2}}}$ \\  \hline extra gauge boson $Z'$
\cite{Langacker:2008yv,Rizzo:2006nw,Leike:1998wr,Hewett:1988xc} &
${\displaystyle{g_\alpha'{}^e\,g_\beta'{}^f\,\chi_{Z'}}}$ \\
\hline AGC ($f=\ell$) \cite{Gounaris:1997ft} &
${\displaystyle{\Delta_{\rm LL}=s\left(\frac{\tilde
f_{DW}}{2s_W^2}+ \frac{2 \tilde f_{DB}}{c_W^2}\right)}}$,
${\displaystyle{\frac{\Delta_{\rm RR}}{2}=\Delta_{\rm
LR}=\Delta_{\rm RL} =s\frac{4 \tilde f_{DB}}{c_W^2}}}$  \\  \hline
TeV-scale extra dim. \cite{Pasztor:2001hc,Cheung:2001mq} &
$-{\displaystyle{(Q_eQ_f+g_\alpha^e\,g_\beta^f)\,\frac{\pi^2\,s}{3\,M_C^2}}}$
\\  \hline
ADD model \cite{ArkaniHamed:1998rs,Hewett:1998sn} &
${\displaystyle{\Delta_{\rm LL}=\Delta_{\rm RR}=f_G\,(1-2\,z)}}$,
${\displaystyle{\Delta_{\rm LR}=\Delta_{\rm RL}=-f_G\,(1+2\,z)}}$
\\  \hline
$R$-parity violating SUSY \cite{Kalinowski:1997bc,Rizzo:1998vf} &
\multirow{2}{*}{ ${\displaystyle{\Delta_{\rm LL}=\Delta_{\rm RR}=0}}$,
${\displaystyle{\Delta_{\rm LR}=\Delta_{\rm RL}=
\frac{1}{2}\,C_{\tilde\nu}^t\, \chi_{\tilde{\nu}}^{t} }}$}
\\
($\tilde{\nu}$  exchange in $t$-channel)
 &
\\  \hline
\end{tabular}
\end{center}
\label{CI}
\caption{ Parametrization of the $\Delta_{\alpha\beta}$ functions
in different NP models ($\alpha,\beta={\rm L,R}$). For the explanation of notation see text. }
\end{table}
In Table~1 $\Lambda_{\alpha\beta}$ denote compositeness scales;
$\chi_{Z'}$ and $\chi_{\tilde{\nu}}^t$ parametrize the $Z'$ and
sneutrino propagators defined analogously to Eq.~(\ref{Eq:chi}),
with superscript $t$ referring to the $t$-channel, e.g.,
$\chi_{\tilde{\nu}}^t=s/(t-M_{\tilde\nu}^2)$, where $M_{\tilde\nu}$
is the sneutrino mass. For the $t$-channel  $\tilde{\nu}$
sneutrino exchange
$C_{\tilde\nu}^t=\lambda^2/{4\pi\alpha_\text{em}}$ with $\lambda$
being the relevant Yukawa coupling. $g_\alpha'{}^f$ parametrizes
the $Z'$ couplings to the $f$ current of chirality $\alpha$.
Furthermore, $\tilde f_{DW}$ and $\tilde f_{DB}$ are related to
$f_{DW}$ and $f_{DB}$ of ref.~\cite{Gounaris:1997ft} by $\tilde
f=f/m_t^2$ ($f_{DW}$ and $f_{DB}$ parametrize new-physics effects
associated with the SU(2) and hypercharge currents, respectively);
$M_C$ is the compactification scale; $f_G=\pm
s^2/(4\pi\alpha_\text{em} M_H^4)$ parametrizes the strength
associated with massive graviton exchange with $M_H$ the cut-off
scale in the KK graviton tower sum.

The doubly polarized total cross section can
be obtained from Eq.~(\ref{cross}) after integration over $z$ within the interval $-1\leq z\leq 1$.
In the limit of $s$, $t$ small compared to the CI mass scales,  the result takes the form
\begin{equation}
\sigma^{\rm CI} = {\sigma}_+^{\rm CI} + {\sigma}_- ^{\rm
CI}=\frac{1}{4}\, \left[(1-P^-)(1+{P^+})\,(\sigma_{\rm LL}^{\rm
CI}+ \sigma_{\rm LR}^{\rm CI}) +(1+P^-)(1-{P^+})\,(\sigma_{\rm
RR}^{\rm CI}+\sigma_{\rm RL}^{\rm CI})\right].
 \label{crosstot}
\end{equation}
It is clear that the formula in the SM has the same form
where  one should replace the superscript ${\rm CI}\to{\rm SM}$ in
Eq.~(\ref{crosstot}).

Since the   $\tilde \nu$ exchanged in the $s$-channel does not interfere with the $s$-channel SM $\gamma$ and $Z$ exchanges, the differential cross section with  both electron
and positron beams polarized can be written as \cite{Rizzo:1998vf,
Osland:2003fn}
\begin{equation}
\frac{\dd\sigma^{\tilde\nu}}{\dd z} =\frac{3}{8} \left[(1+z)^2
{\sigma}_+^{\rm SM} +(1-z)^2 {\sigma}_-^{\rm SM}+
2\,\frac{1+P^-{P^+}}{2}\,(\sigma_{\rm RL}^{\tilde\nu}+\sigma_{\rm
LR}^{\tilde\nu}) \right]. \label{snucross}
\end{equation}
Here, $\sigma_{\rm RL}^{\tilde\nu}(=\sigma_{\rm
LR}^{\tilde\nu})=\sigma_{\rm pt}\, \vert{\cal M}^{\tilde\nu}_{\rm
RL}\vert^2$, ${\cal M}^{\tilde\nu}_{\rm RL}={\cal
M}^{\tilde\nu}_{\rm LR}=\frac{1}{2}\,C_{\tilde\nu}^s\chi_{\tilde\nu}^s$, and
$C_{\tilde{\nu}}^s$ and $\chi_{\tilde{\nu}}^s$ denote  the product of the $R$-parity violating couplings
and the propagator of the exchanged sneutrino. For the $s$-channel  $\tilde{\nu}_\tau$
sneutrino exchange they read
\begin{equation} \label{Eq:Ctilde}
C_{\tilde\nu}^s\, \chi_{\tilde{\nu}}^s =\frac{\lambda_{131}\lambda_{232}}{4\pi\alpha_\text{em}}\, \frac{s}{s-M^2_{\tilde{\nu}_\tau}+iM_{\tilde{\nu}_\tau}\Gamma_{\tilde{\nu}_\tau}}
\end{equation}
Below we will use  the abbreviation $\lambda^2=\lambda_{131}\lambda_{232}$.

As seen from
Eq.~(\ref{snucross}) the  polarized differential cross section
 picks up a $z$-independent term in addition to the SM part. The corresponding total cross
 section can be written as

\begin{eqnarray}
\sigma^{\tilde\nu} &=&\frac{1}{4}\,(1-P^-)(1+{P^+})\,(\sigma_{\rm
LL}^{\rm SM}+ \sigma_{\rm LR}^{\rm SM})
+\frac{1}{4}\,(1+P^-)(1-{P^+})\,(\sigma_{\rm RR}^{\rm SM}
+\sigma_{\rm RL}^{\rm SM})\nonumber \\
&+& \frac{3}{2}\,\frac{1+P^-{P^+}}{2}\,(\sigma_{\rm
RL}^{\tilde\nu}+\sigma_{\rm LR}^{\tilde\nu}). \label{totsnu}
\end{eqnarray}

It is possible to uniquely identify the effect of the $s$-channel sneutrino exchange exploiting the
double beam polarization asymmetry defined as
\cite{Rizzo:1998vf,Osland:2003fn}
\begin{equation}
A_{\rm double}=\frac
{\sigma(P_1,-P_2)+\sigma(-P_1,P_2)-\sigma(P_1,P_2)-\sigma(-P_1,-P_2)}
{\sigma(P_1,-P_2)+\sigma(-P_1,P_2)+\sigma(P_1,P_2)+\sigma(-P_1,-P_2)},
\label{double}
\end{equation}
where $P_1=\vert{P^-}\vert$, $P_2=\vert{P^+}\vert$.  It can easily be checked for  the whole set of contact interactions listed in Table~1, with the exception of the $s$-channel sneutrino exchange, that  from
(\ref{crosstot}) and (\ref{double}) one finds
\begin{equation}
A_{\rm double}^{\rm SM}=A_{\rm double}^{\rm CI}=P_1P_2=0.48,
\label{ACI}
\end{equation}
where the numerical value corresponds to electron and positron
degrees of polarization: $P_1=0.8$, $P_2=0.6$. This is because these
contact interactions contribute to the same amplitudes as shown in
(\ref{amplit}). Eq.~(\ref{ACI}) demonstrates that $A_{\rm
double}^{\rm SM}$ and $A_{\rm double}^{\rm CI}$ are
indistinguishable for any values of the contact interaction parameters,
$\Delta_{\alpha\beta}$, i.e. $\Delta A_{\rm double}= A_{\rm
double}^{\rm CI}-A_{\rm double}^{\rm SM}=0$.

\begin{figure}[tbh!] 
\centerline{
\includegraphics[width=0.49\textwidth]{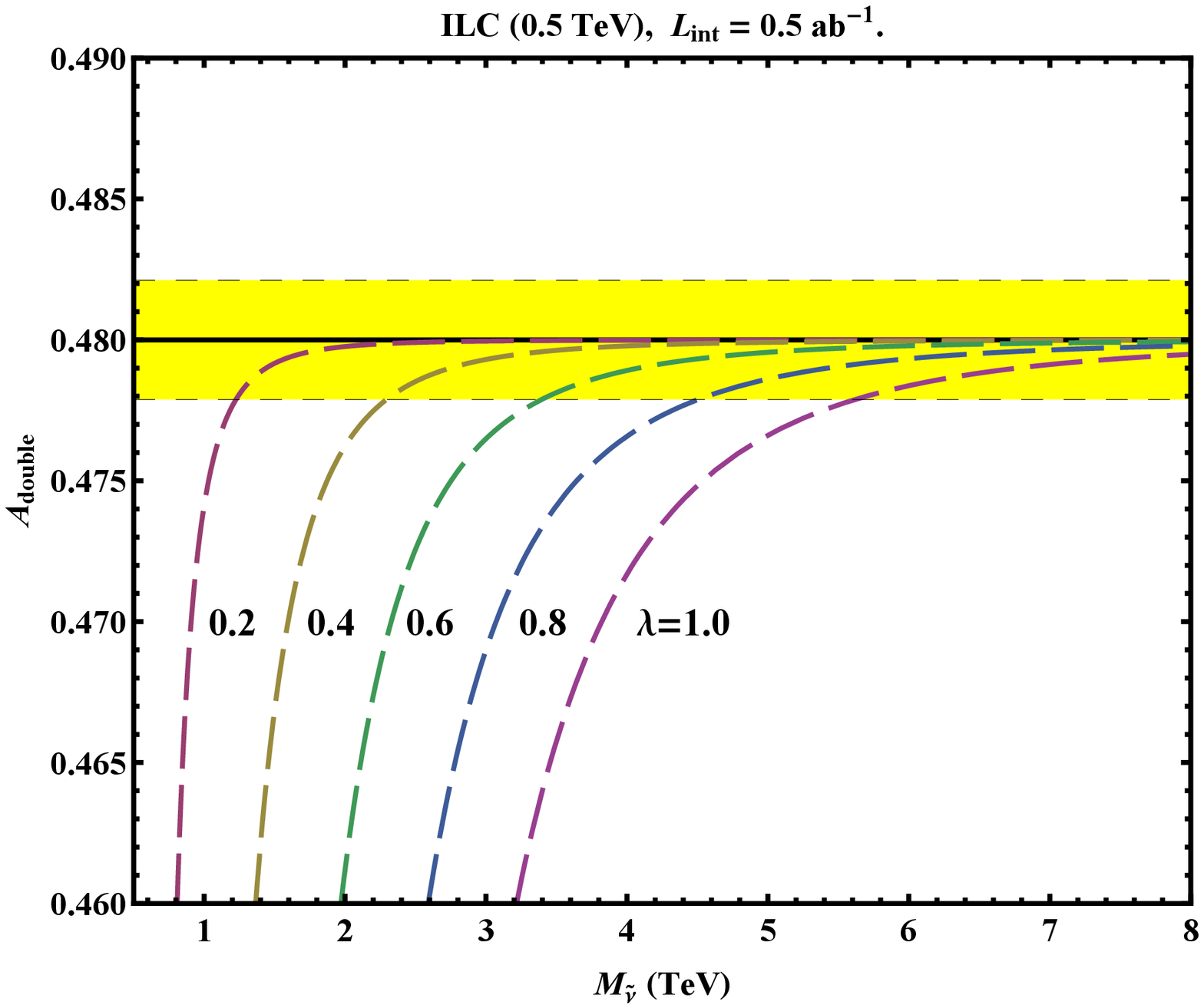}
\includegraphics[width=0.49\textwidth]{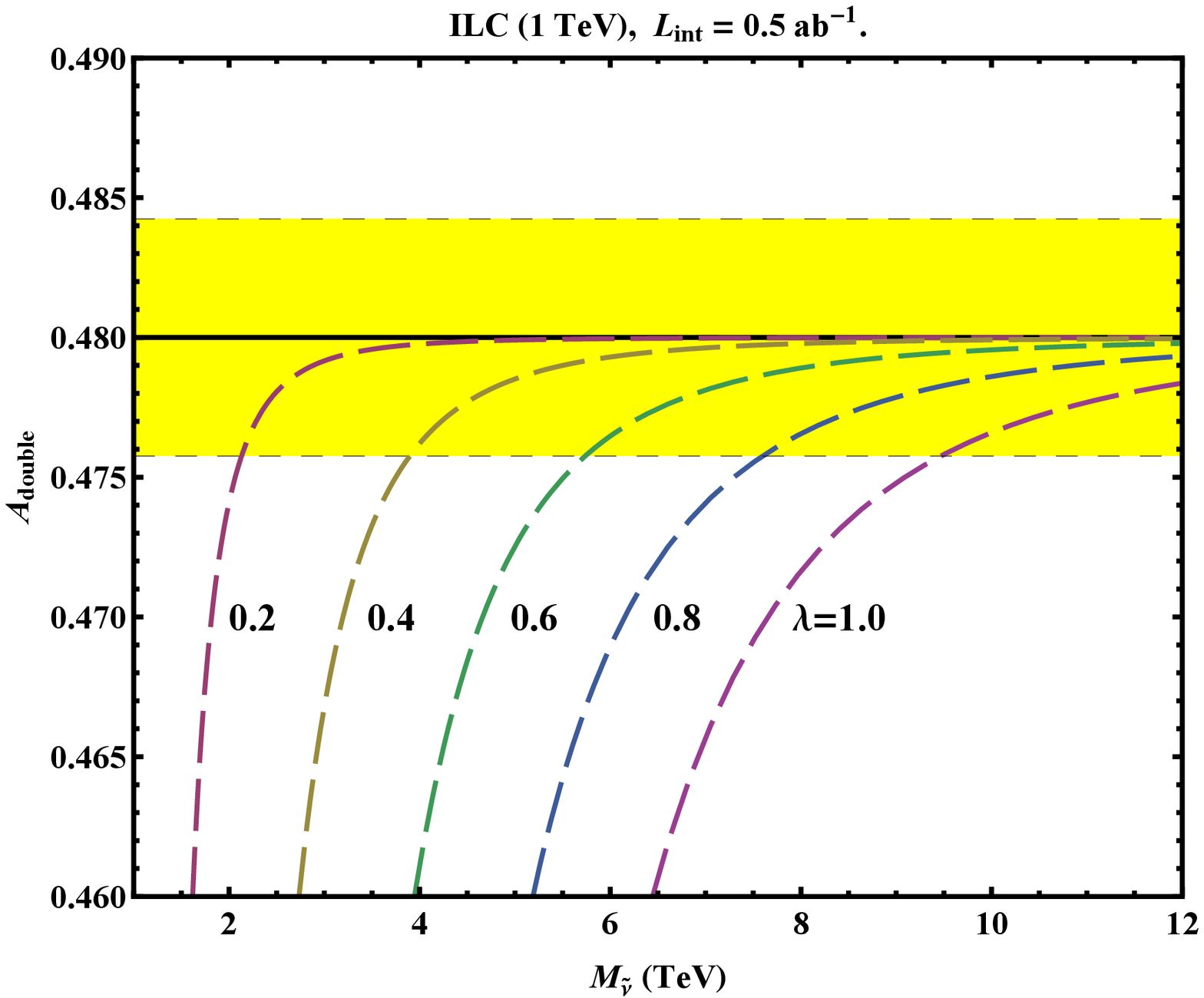}
}
\caption{\label{fig1}  Double beam polarization asymmetry $A_{\rm
double}^{\tilde\nu}$ as a function of sneutrino mass
$M_{\tilde\nu}$ for different choices of $\lambda$ (dashed lines)
at the ILC with $\sqrt{s}=0.5$ TeV (left panel) and $\sqrt{s}=1.0$
TeV (right panel), $\Lumint=0.5~\text{ab}^{-1}$. From left to right,
$\lambda$ varies from 0.2 to 1.0 in steps of 0.2. The solid
horizontal line corresponds to $A_{\rm double}^{\rm SM}=
A_{\rm double}^{\rm CI}$. The yellow bands indicate the expected uncertainty in the SM case.}
\end{figure}

On the contrary, the $\tilde\nu$ exchange in the $s$-channel will force
this observable to a smaller value, $\Delta A_{\rm double}=
A_{\rm double}^{\tilde\nu}-A_{\rm double}^{\rm SM} \propto
-P_1P_2\,|C^s_{\tilde\nu}\chi^s_{\tilde\nu}|^2<0$. The value of $A_{\rm
double}$  below $P_1P_2$ can provide a signature of scalar
exchange in the $s$-channel. All those features in the $A_{\rm
double}$ behavior are shown in Fig.~\ref{fig1}.

The non-zero value of the $\lambda_{131}$ coupling implies that the Bhabha scattering process will receive $\tilde{\nu}_\tau$ contributions from both the $s$- and $t$-channel exchanges.  The differential cross section can
be written in this case as
\begin{align}
  \frac{\dd\sigma^{\tilde\nu}}{\dd z}
  =
      \frac{\pi \alpha_\text{em}^2}{8 s}
       \bigl[
       & (1+z)^2
          \{
(1-P^-)(1+{P^+}) \, |f^{s}_{LL}+f^{t}_{LL}|^2
+ (1+P^-)(1-{P^+}) \, |f^{s}_{RR}+f^{t}_{RR}|^2
          \}
         \nonumber \\
  + &(1-z)^2
           \{
             {(1-P^-)(1+{P^+})} \, |f^{s}_{LR}|^2 +
             {(1+P^-)(1-{P^+})} \, |f^{s}_{RL}|^2
          \} \nonumber \\
          &+ 4 \, (1 + P^-{P^+}) \{ |f^{t}_{LR}|^2 + |f^{t}_{RL}|^2 \}
          \bigr]
          \label{bhabhadiff}
\end{align}
where\footnote{Note that Ref.~\cite{Kalinowski:1997bc}, for example, uses a different convention for the chirality of the final state current.}
\begin{alignat}{2}
   f^{s}_{LL} &= 1 + (g^{e}_{L})^2 \chi_{Z},        &\quad
   f^{s}_{RR} &= 1 + (g^{e}_{R})^2 \chi_{Z},        \nonumber \\
   f^{s}_{LR} &= 1 +  g^{e}_{L} g^{e}_{R} \chi_{Z} + \frac{1}{2} C_{\tilde{\nu}} \chi_{\tilde{\nu}}^{t},&\quad
   f^{s}_{RL} &= 1 +  g^{e}_{R} g^{e}_{L} \chi_{Z} + \frac{1}{2} C_{\tilde{\nu}} \chi_{\tilde{\nu}}^{t}, \nonumber \\
   f^{t}_{LL} &= \frac{s}{t} +  (g^{e}_{L})^2 \chi_{Z}^{t}, &\quad
   f^{t}_{RR} &= \frac{s}{t} +  (g^{e}_{R})^2 \chi_{Z}^{t}, \nonumber \\
   f^{t}_{LR} &= \frac{s}{t} +  g^{e}_{L} g^{e}_{R} \chi_{Z}^{t} + \frac{1}{2} C_{\tilde{\nu}} \chi_{\tilde{\nu}}^s, &\quad
   f^{t}_{RL} &= \frac{s}{t} +  g^{e}_{R} g^{e}_{L} \chi_{Z}^{t} + \frac{1}{2}  C_{\tilde{\nu}} \chi_{\tilde{\nu}}^s,
\label{bhabhaamplit}
\end{alignat}
where $\chi^t_i=s/(t-M^2_i)$
Note that we use the same notation as in Eq.~(\ref{Eq:Ctilde}) for the reduced sneutrino coupling $C_{\tilde{\nu}}$. However, since now the same lepton generation is present in the initial and final states, consequently in Eq.~(\ref{bhabhaamplit}) we have
\begin{equation}
C_{\tilde{\nu}}=\frac{\lambda_{131}^2}{4\pi\alpha_\text{em}}
\end{equation}
for both $s$- and $t$-channel sneutrino exchanges.
\section{Numerical analysis}\label{sec:numeric}

In the  numerical analysis, cross sections are evaluated
including initial- and final-state radiation by means of the
program ZFITTER \cite{Bardin:1999yd}, together
with ZEFIT \cite{Sabine}, with $m_{\rm
top}=175$~GeV and $m_H=125$~GeV. One-loop SM electroweak
corrections are accounted for by improved Born amplitudes
\cite{Consoli:1989pc}, such that the forms of the previous
formulae remain the same. Concerning initial-state radiation, a
cut on the energy of the emitted photon $\Delta=E_\gamma/E_{\rm
beam}=0.9$ is applied in order to avoid the radiative return to
the $Z$ peak and enhance the signal originating from the
nonstandard physics contribution \cite{Djouadi:1991sx}.

As numerical inputs, we shall assume  the identification
efficiencies of $\epsilon=95\%$ for $\mu^+\mu^-$ final states, integrated
luminosity of $\Lumint=0.5~\text{ab}^{-1}$  with uncertainty
$\delta\Lumint/\Lumint=0.5\%$, and a fiducial experimental angular
range $|\cos\theta|\le 0.99$. Also, regarding electron and
positron degrees of polarization, we shall consider the following
values: $ P^-=\pm 0.8$; $P^+ =\pm 0.6$, with $\delta
P^-/P^-=\delta{P^+}/{P^+}=0.5\%$.

\begin{figure}[tbh!] %
\centerline{
\includegraphics[width=0.49\textwidth]{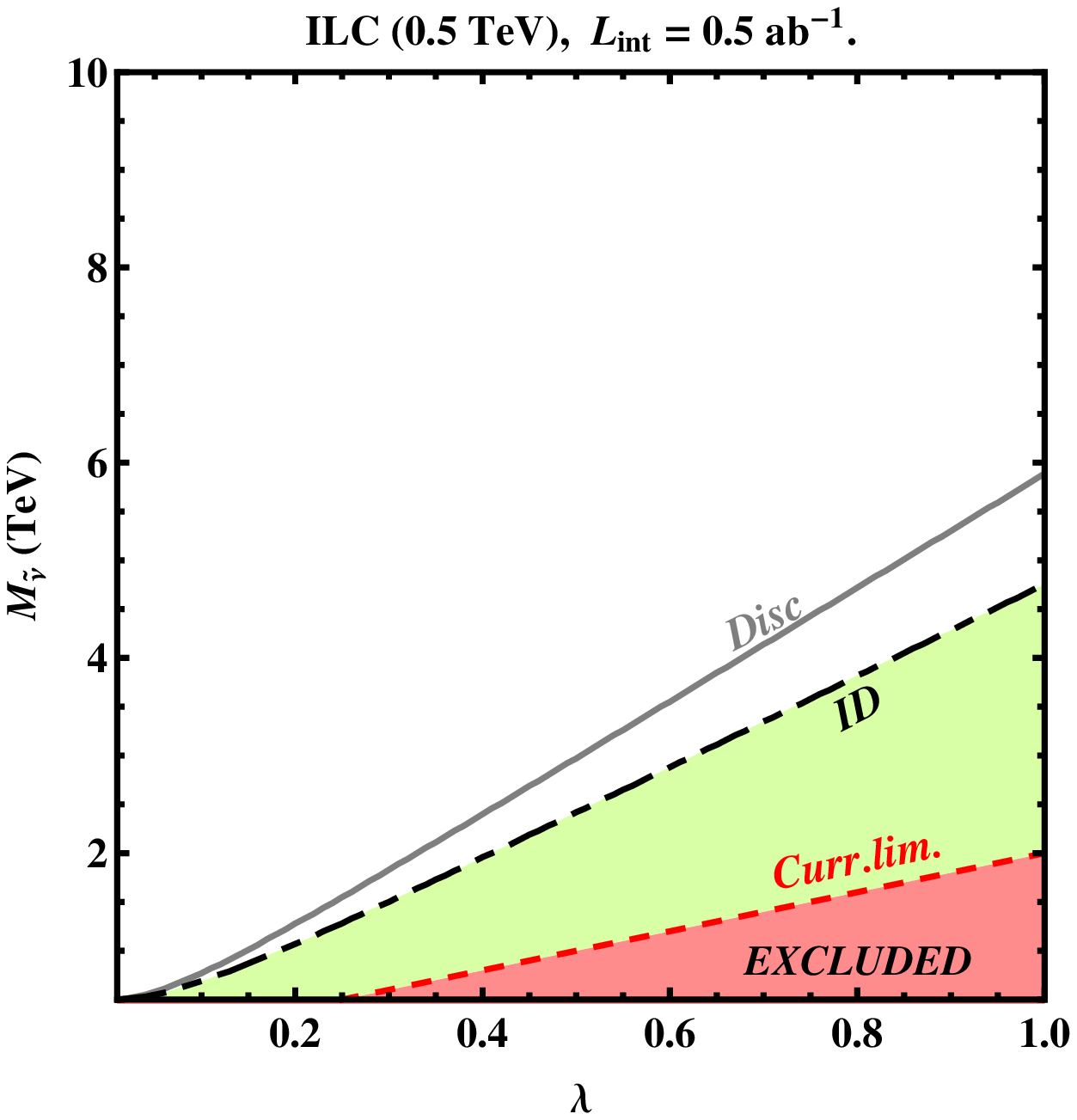}
\includegraphics[width=0.49\textwidth]{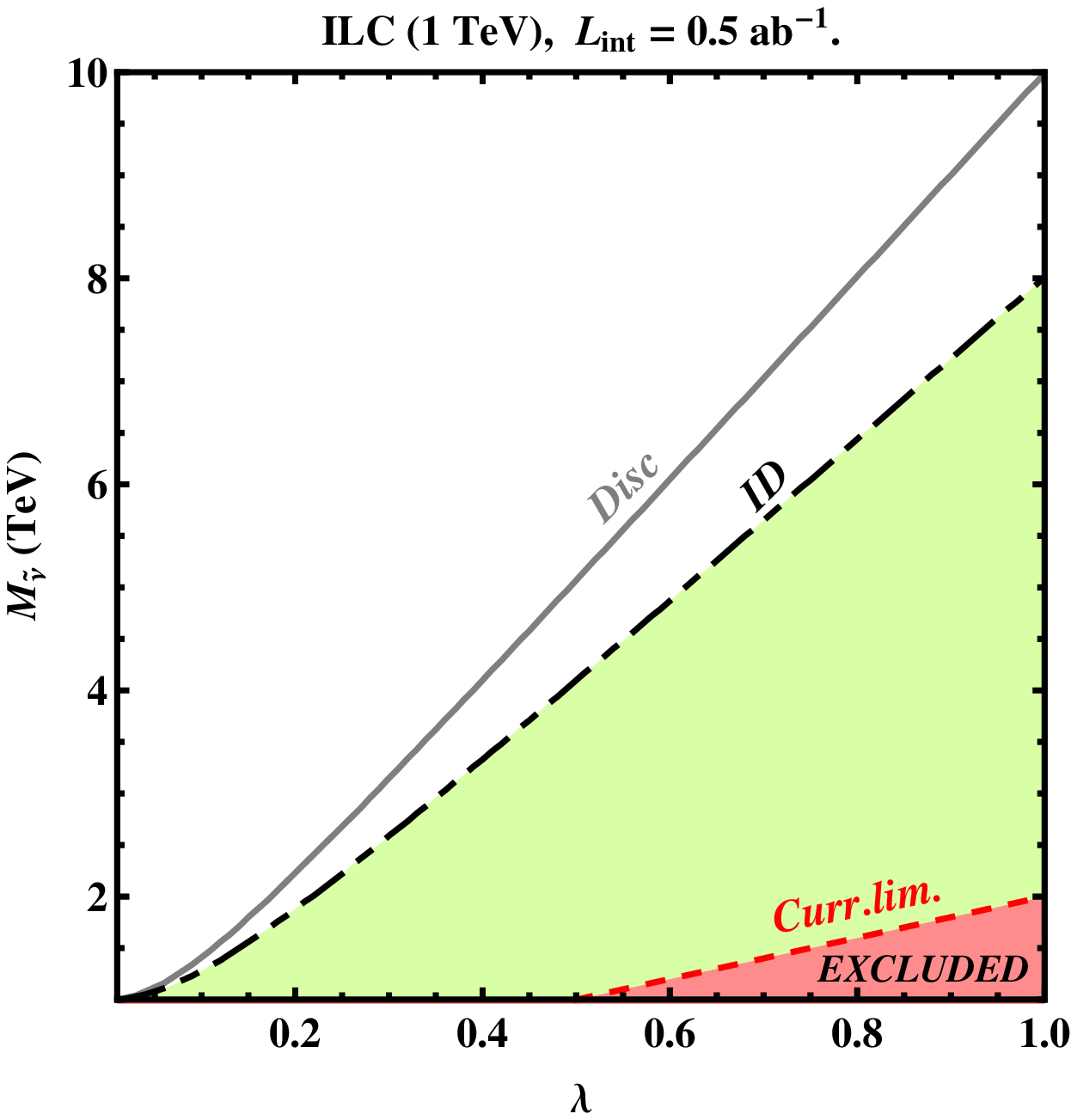}}
\caption{\label{fig2} Discovery and identification reaches on
sneutrino mass $M_{\tilde\nu}$ (95\% C.L.) as a function of
$\lambda$ for the process $e^+e^-\to\mu^+\mu^-$ at the ILC with $\sqrt{s}=0.5$ TeV (left panel) and
$\sqrt{s}=1.0$ TeV (right panel), $\Lumint=0.5~\text{ab}^{-1}$.
For comparison, current limits from low energy data are
also displayed. }
\end{figure}

Discovery and identification reaches on the sneutrino mass
$M_{\tilde\nu}$ (95\% C.L.) plotted in Fig.~\ref{fig2} are
obtained from conventional $\chi^2$ analysis. The discovery limit
(Disc) is obtained from a combined analysis of the polarized
differential cross sections, $d\sigma/dz$, in 10 equal-size
$z$-bins in the range $[-0.99,0.99]$, with beam polarizations of
the same sign, $(P^-,{P^+})=(+0.8,+0.6); (-0.8,-0.6)$. This
procedure provides the best sensitivity to sneutrino parameters,
whereas the identification reach (ID) is derived from $A_{\rm
double}$.
 In the latter case the $\chi^2$ function is constructed as follows:
 $\chi^2=(\Delta A_{\rm double}/\delta A_{\rm double})^2$ where $\delta A_{\rm
 double}$ is the expected experimental uncertainty accounting for both statistical and
 systematic components.

For comparison, current limits from low-energy data are also shown
\cite{Kao:2009fg,Bhattacharyya:2011zv}. From Fig.~\ref{fig2} one
can see  that identification of sneutrino exchange effects in the
$s$-channel with $A_{\rm double}$ is feasible in the region of
parameter and mass space far beyond the current limits.

As was demonstrated in Ref.~\cite{Osland:2010yg} the resonant
$s$-channel production of sneutrino $\tilde \nu$ with their
subsequent decay into purely leptonic final states via $R$-parity
violating couplings can be observed over a wide range of
parameters (couplings and masses) in hadronic collisions
(\ref{proc_DY}). This process provides a clean and powerful probe
of $R$-parity violating supersymmetric parameter space and the
corresponding LHC search reaches in the parameter plane spanned by
the sneutrino mass and the $R$-parity-violating coupling were
obtained there. Specifically, in the dilepton process
(\ref{proc_DY}) of interest here, a spin-0 sneutrino can be
exchanged through the subprocess $d{\bar d} \to{\tilde\nu}\to
l^+l^-$ and manifest itself as a peak in the dilepton invariant mass
distribution and also with a flat angular distribution. The cross
section is proportional to the $R$-parity violating product
$X=(\lambda^\prime)^2B_l$ where $B_l$ is the sneutrino leptonic
branching ratio and $\lambda^\prime$ the relevant sneutrino
coupling to the $d{\bar d}$ quarks. The experimental 95\% CL lower
limits on $M_{\tilde\nu}$ range from 397 GeV (for $X=10^{-4}$) to
866 GeV (for $X=10^{-2}$) \cite{Aaltonen:2008ah}.

If this signature is observed, the leptonic center-edge integrated
asymmetry \cite{Dvergsnes:2004tw} can be successfully used to
distinguish slepton resonances from those associated with new
spin-1 $Z^\prime$ gauge bosons and the Randall-Sundrum graviton
resonance (spin-2). Once large integrated luminosities of order
$\sim 100$ fb$^{-1}$ are obtained at the LHC, these new scalar
resonances should be visible for masses as large as $\sim 1.5-5.5$
TeV depending on the specific details of the model (couplings and
leptonic branching ratios). Accordingly, the analysis performed in
\cite{Osland:2010yg} indicates that the identification of the
sneutrino against the RS graviton and $Z'$ bosons by center-edge
asymmetry is possible at the LHC for $M_{\tilde\nu}\leq$ 4.5 TeV
for $X$ in the range of $10^{-5}<X<10^{-1}$.

As mentioned above, future $e^+e^-$ colliders operating in the TeV
energy range can indirectly probe for new  physics effects  by
exploring contact-interaction-like deviations from the cross
sections and asymmetries predicted by the SM. For luminosity
expected at ILC, $\sim 0.5$ ab$^{-1}$, and with both electron and
positron beams polarized, from Fig.~\ref{fig2} we see that this
implies that the parameter space region $\lambda/M_{\tilde\nu}>$
0.17 (0.10) ($M_{\tilde\nu}$ in TeV unit) would certainly be
probed at $\sqrt{s}=0.5$ (1) TeV by such measurements while
identification parameter space populates the region 0.21 (0.13)
$<\lambda/M_{\tilde\nu}<$0.5.

For Bhabha
scattering, the angular range $\vert\cos\theta\vert< 0.90$ is
divided into nine equal-size bins. We combine the cross sections
with the following initial electron and positron longitudinal
polarizations: $(P^-,P^+)=(\vert P^-\vert,-\vert P^+\vert)$;
$(-\vert P^-\vert,\vert P^+\vert$; $(\vert P^-\vert,\vert
P^+\vert)$; $(-\vert P^-\vert,-\vert P^+\vert)$. The assumed
reconstruction efficiencies, that determine the expected
statistical uncertainties, are 100\% for $e^+e^-$ final pairs.
Concerning the ${\cal O}(\alpha_\text{em})$ QED corrections, the
(numerically dominant) effects from initial-state radiation for
Bhabha scattering  are again accounted for by a structure function
approach including both hard and soft photon emission
\cite{nicrosini}, and by a flux factor method
\cite{physicsatlep2}, respectively.

\begin{figure}[tbh!] %
\centerline{
\includegraphics[width=0.45\textwidth]{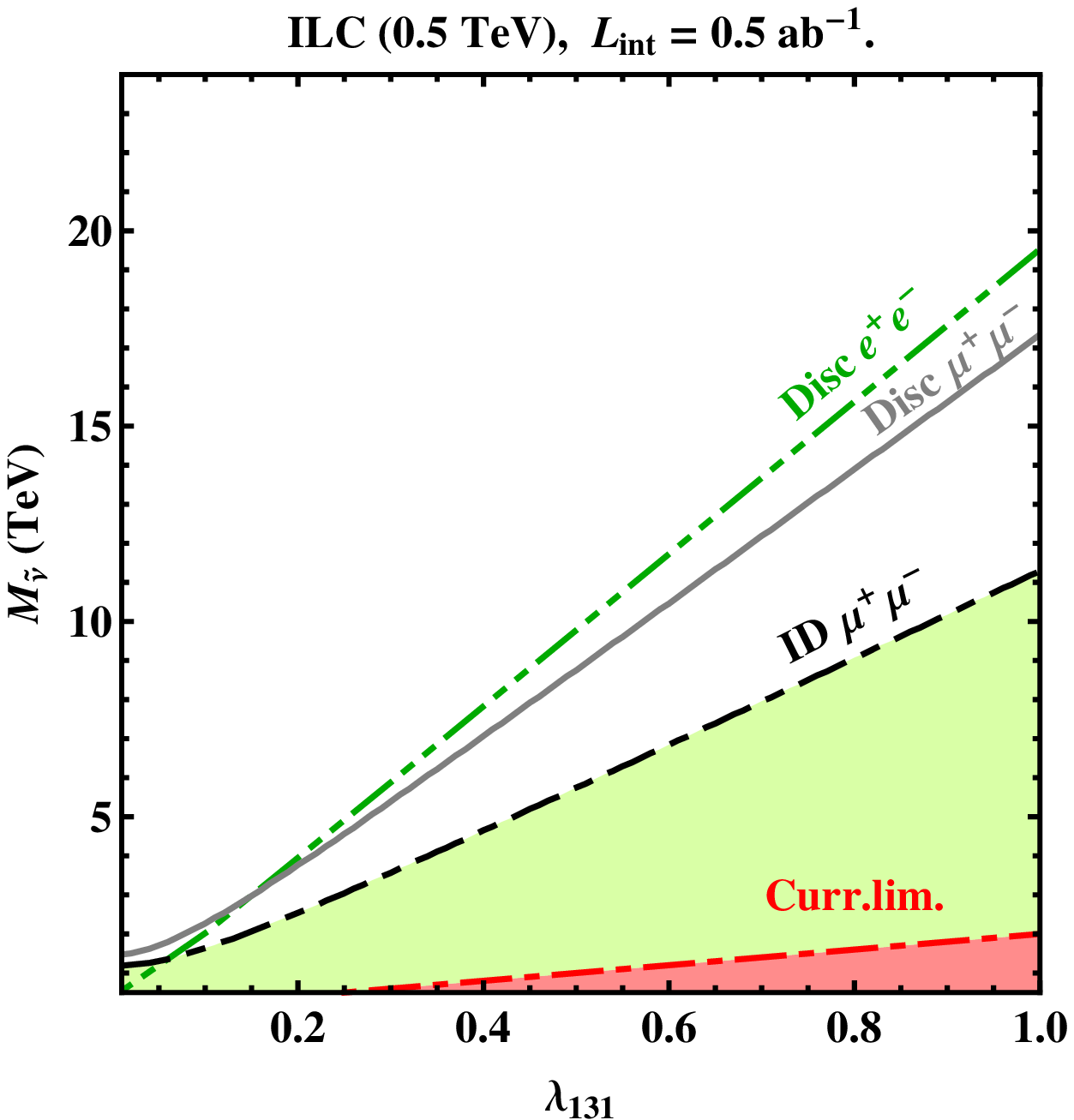}
\includegraphics[width=0.45\textwidth]{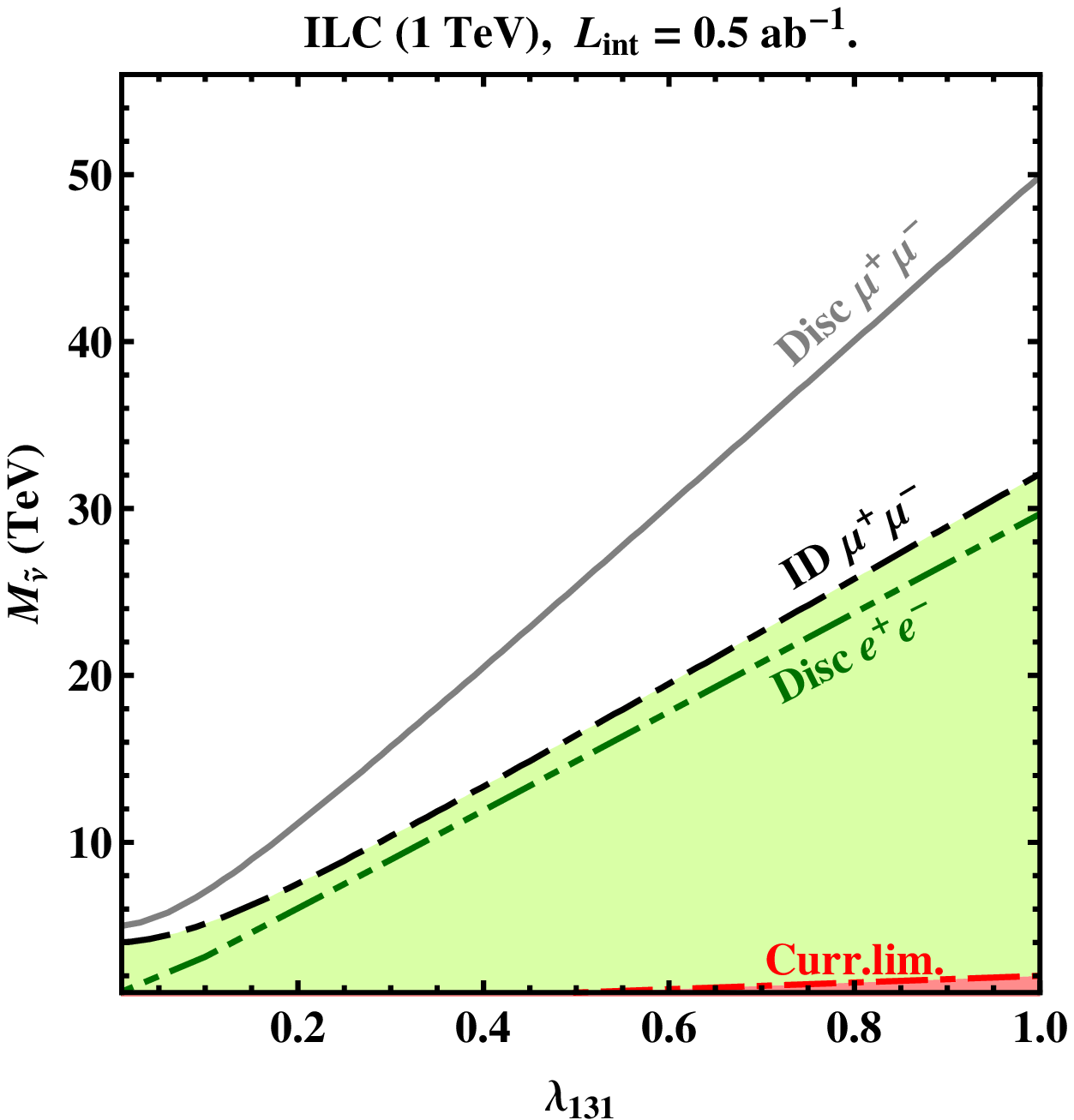}
}
\caption{\label{fig3}  Discovery reach on sneutrino mass (95\%
C.L.) in Bhabha scattering as a function of $\lambda_{131}$ at $\sqrt{s}=0.5~\text{TeV}$ (left panel) and 1~TeV (right panel), for $\Lumint=0.5~\text{ab}^{-1}$. For
comparison, discovery reach on $M_{\tilde\nu}$ in muon pair production is also
depicted for $\lambda_{232}=0.5\times M_{\tilde\nu}/\text{TeV}$.}
\end{figure}

One can parametrize the bounds depicted in
Fig.~\ref{fig2} (in the plane ($M_{\tilde\nu}$, $\lambda$)) approximately
as a straight line,
$M_{\tilde\nu}=k_\mu\lambda$ ($M_{\tilde\nu}$ is taken in TeV
units), $\lambda=\sqrt{\lambda_{131}\cdot\lambda_{232}}$ and
$k_\mu$ is the slope of the these lines for the process $e^+e^-\to\mu^+\mu^-$.
For instance, for the discovery reach we have $k_\mu\approx5.9$
(10) for $\sqrt s=0.5$ (1) TeV. In order to convert the bounds
shown in Fig.~\ref{fig2} into limits on $M_{\tilde\nu}$   vs
$\lambda_{131}$ one should fix $\lambda_{232}$. For that purpose
one can take the (mass dependent) current limit on that Yukawa
coupling $\lambda_{232}$ represented as
$\lambda_{232}/M_{\tilde\nu}=0.5$. From these formulae one finds:
$M_{\tilde\nu}<(k_\mu^2/2)\,\lambda_{131}$. These
areas which can be explored in the muon pair production process are shown in
Fig.~\ref{fig3}. In contrast to the limits shown in
Fig.~\ref{fig2} as curves, limits on $M_{\tilde\nu}$ vs
$\lambda_{131}$ for both the discovery and the identification are represented in Fig.~\ref{fig3} as areas
constrained by the line for the current limit,
$M_{\tilde\nu}=2\lambda_{131}$, and the lines for the upper bounds,
$M_{\tilde\nu}=(k_\mu^2/2)\,\lambda_{131}$.

In contrast to muon pair production, identification of the sneutrino exchange effects by means
of Bhabha scattering is impossible because CI and sneutrino give
rise to the same helicity amplitudes as clearly seen from
(\ref{bhabhadiff}) and (\ref{bhabhaamplit}) \cite{Pankov:2005kd}. Therefore only the discovery reach for the Bhaba process is shown in the figure.

\section{Concluding remarks}
\label{sect:concl}

In this note we have studied how uniquely
identify the indirect (propagator) effects of spin-0 sneutrino
predicted by supersymmetric theories with $R$-parity violation,
against other new physics scenarios in high energy $e^+e^-$
annihilation into lepton-pairs at the ILC. The competitive models
are the interactions based on gravity in large and in TeV-scale
extra dimensions, anomalous gauge couplings, extra $Z'$ bosons,
and the compositeness-inspired four-fermion contact interactions.
All those kinds of new physics can lead to qualitatively similar
modifications of SM cross sections, angular distributions and
various asymmetries, but they differ in detail. To evaluate the
identification reach on the sneutrino exchange signature, we
develop a technique based on a double polarization asymmetry
formed by polarizing both beams in the initial state, that is
particularly suitable to directly test for such $s$-channel
sneutrino exchange effects in the data analysis. We show that  the
availability of both beams being polarized, plays a crucial
r{\^o}le in identifying that new physics scenario, as the commonly
considered asymmetry, $A_\text{LR}$, formed when only a single
beam is polarized, was shown not to be useful for the purpose of
sneutrino identification.


\section*{Acknowledgements}

It is a pleasure to thank G. Moortgat-Pick for helpful discussions.  This
research has been partially supported by the Abdus Salam ICTP (TRIL
Programme and Associates Scheme) and the Collaborative Research Center
SFB676/1-2006 of the DFG at the Department of Physics, University of
Hamburg. JK was partially supported by the Polish Ministry of Science and
Higher Education Grant N N202 230337. The work of AVT has been partially
supported by the INFN ``Fondo Affari Internazionali''. The work of PO has
been supported by the Research Council of
Norway.




\begin{thebibliography}{99}

\bibitem{Eichten:1983hw} 
  E.~Eichten, K.~D.~Lane and M.~E.~Peskin,
  Phys.\ Rev.\ Lett.\  {\bf 50}, 811 (1983); \\
  R.~R\"uckl,
  Phys.\ Lett.\ B {\bf 129}, 363 (1983).

\bibitem{Langacker:2008yv}
  P.~Langacker,
  Rev.\ Mod.\ Phys.\  {\bf 81}, 1199-1228 (2009).
  [arXiv:0801.1345 [hep-ph]].

\bibitem{Rizzo:2006nw}
  T.~G.~Rizzo,
    [hep-ph/0610104].

\bibitem{Leike:1998wr}
  A.~Leike,
  Phys.\ Rept.\  {\bf 317}, 143-250 (1999).
  [hep-ph/9805494].

\bibitem{Hewett:1988xc}
  J.~L.~Hewett, T.~G.~Rizzo,
  Phys.\ Rept.\  {\bf 183}, 193 (1989).

\bibitem{Buchmuller:1986zs} 
  W.~B\"uchmuller, R.~R\"uckl and D.~Wyler,
  Phys.\ Lett.\ B {\bf 191}, 442 (1987)
  [Erratum-ibid.\ B {\bf 448}, 320 (1999)]; \\
  G.~Altarelli, J.~R.~Ellis, G.~F.~Giudice, S.~Lola and M.~L.~Mangano,
  Nucl.\ Phys.\ B {\bf 506}, 3 (1997)
  [hep-ph/9703276]; \\
  R.~Casalbuoni, S.~De Curtis, D.~Dominici and R.~Gatto,
  Phys.\ Lett.\ B {\bf 460}, 135 (1999)
  [hep-ph/9905568]; \\
  V.~D.~Barger and K.~-m.~Cheung,
  Phys.\ Lett.\ B {\bf 480}, 149 (2000)
  [hep-ph/0002259].

\bibitem{Kalinowski:1997bc} 
  J.~Kalinowski, R.~R\"uckl, H.~Spiesberger and P.~M.~Zerwas,
  Phys.\ Lett.\ B {\bf 406}, 314 (1997)
  [hep-ph/9703436]; 
  Phys.\ Lett.\ B {\bf 414}, 297 (1997)
  [hep-ph/9708272].

\bibitem{Rizzo:1998vf} 
  T.~G.~Rizzo,
  Phys.\ Rev.\ D {\bf 59}, 113004 (1999)
  [hep-ph/9811440].

\bibitem{Cuypers:1996ia} 
For a review, see
  F.~Cuypers and S.~Davidson,
  Eur.\ Phys.\ J.\ C {\bf 2}, 503 (1998)
  [hep-ph/9609487],
and references therein.

\bibitem{Gounaris:1997ft} 
  G.~J.~Gounaris, D.~T.~Papadamou and F.~M.~Renard,
  Phys.\ Rev.\ D {\bf 56}, 3970 (1997)
  [hep-ph/9703281].

\bibitem{ArkaniHamed:1998rs} 
  N.~Arkani-Hamed, S.~Dimopoulos and G.~R.~Dvali,
  Phys.\ Lett.\ B {\bf 429}, 263 (1998)
  [hep-ph/9803315]; 
  Phys.\ Rev.\ D {\bf 59}, 086004 (1999)
  [hep-ph/9807344]; \\
  L.~Randall and R.~Sundrum,
  Phys.\ Rev.\ Lett.\  {\bf 83}, 3370 (1999)
  [hep-ph/9905221]; \\
  I.~Antoniadis, N.~Arkani-Hamed, S.~Dimopoulos and G.~R.~Dvali,
  Phys.\ Lett.\ B {\bf 436}, 257 (1998)
  [hep-ph/9804398].

\bibitem{Giudice:1998ck} 
  G.~F.~Giudice, R.~Rattazzi and J.~D.~Wells,
  Nucl.\ Phys.\ B {\bf 544}, 3 (1999)
  [hep-ph/9811291]; \\
  Nucl.\ Phys.\ B {\bf 630}, 293 (2002)
  [hep-ph/0112161].

\bibitem{Hewett:1998sn} 
  J.~L.~Hewett,
  Phys.\ Rev.\ Lett.\  {\bf 82}, 4765 (1999)
  [hep-ph/9811356]; \\
  T.~Han, J.~D.~Lykken and R.~-J.~Zhang,
  Phys.\ Rev.\ D {\bf 59}, 105006 (1999)
  [hep-ph/9811350]; \\
  T.~G.~Rizzo,
  Phys.\ Rev.\ D {\bf 64}, 095010 (2001)
  [hep-ph/0106336]; \\
  H.~Davoudiasl, J.~L.~Hewett and T.~G.~Rizzo,
  Phys.\ Rev.\ Lett.\  {\bf 84}, 2080 (2000)
  [hep-ph/9909255]; 
  Phys.\ Rev.\ D {\bf 63}, 075004 (2001)
  [hep-ph/0006041]; \\
  E.~A.~Mirabelli, M.~Perelstein and M.~E.~Peskin,
  Phys.\ Rev.\ Lett.\  {\bf 82}, 2236 (1999)
  [hep-ph/9811337]; \\
  S.~Cullen, M.~Perelstein and M.~E.~Peskin,
  Phys.\ Rev.\ D {\bf 62}, 055012 (2000)
  [hep-ph/0001166].

\bibitem{Antoniadis:1993jp}
I.~Antoniadis and K.~Benakli,
Phys.\ Lett.\ B {\bf 326}, 69 (1994) [arXiv:hep-th/9310151];
I.~Antoniadis, K.~Benakli and M.~Quiros,
Phys.\ Lett.\ B {\bf 331}, 313 (1994) [arXiv:hep-ph/9403290].

\bibitem{Pasztor:2001hc}
G.~Pasztor and M.~Perelstein,
in {\it Proc. of the APS/DPF/DPB Summer Study on the Future of
Particle Physics (Snowmass 2001) } ed. N.~Graf,
arXiv:hep-ph/0111471.

\bibitem{Cheung:2001mq} 
  K.-m.~Cheung and G.~L.~Landsberg,
  Phys.\ Rev.\ D {\bf 65}, 076003 (2002)
  [hep-ph/0110346].

\bibitem{Bomark:2011ye}
  N.-E.~Bomark, D.~Choudhury, S.~Lola and P.~Osland,
  JHEP {\bf 1107}, 070 (2011)
  [arXiv:1105.4022 [hep-ph]].

\bibitem{MoortgatPick:2005cw}
  G.~Moortgat-Pick {\it et al.},
  Phys.\ Rept.\  {\bf 460}, 131 (2008)
  [arXiv:hep-ph/0507011].

\bibitem{Osland:2010yg}
  P.~Osland, A.~A.~Pankov, N.~Paver and A.~V.~Tsytrinov,
  Phys.\ Rev.\  D {\bf 82}, 115017 (2010)
  [arXiv:1008.1389 [hep-ph]].

\bibitem{Schrempp:1987zy} 
  B.~Schrempp, F.~Schrempp, N.~Wermes and D.~Zeppenfeld,
  Nucl.\ Phys.\ B {\bf 296}, 1 (1988).

\bibitem{Djouadi:1991sx}
  A.~Djouadi, A.~Leike, T.~Riemann, D.~Schaile and C.~Verzegnassi,
  Z.\ Phys.\  C {\bf 56}, 289 (1992).

\bibitem{Osland:2003fn}
  P.~Osland, A.~A.~Pankov and N.~Paver,
  Phys.\ Rev.\  D {\bf 68}, 015007 (2003)
  [arXiv:hep-ph/0304123].

\bibitem{Bardin:1999yd} 
  D.~Y.~Bardin, P.~Christova, M.~Jack, L.~Kalinovskaya, A.~Olchevski, S.~Riemann and T.~Riemann,
  Comput.\ Phys.\ Commun.\  {\bf 133}, 229 (2001)
  [hep-ph/9908433].

\bibitem{Sabine}
S. Riemann, FORTRAN program ZEFIT, Version 4.2.

\bibitem{Consoli:1989pc}
M.~Consoli, W.~Hollik and F.~Jegerlehner,
CERN-TH-5527-89, presented at the {\it Workshop on Z Physics at LEP}; \\
  G.~Altarelli, R.~Casalbuoni, D.~Dominici, F.~Feruglio and R.~Gatto,
  Nucl.\ Phys.\ B {\bf 342}, 15 (1990).

\bibitem{Kao:2009fg}
  Y.~Kao and T.~Takeuchi,
  arXiv:0910.4980 [hep-ph].

\bibitem{Bhattacharyya:2011zv} 
  G.~Bhattacharyya, H.~Pas and D.~Pidt,
  Phys.\ Rev.\ D {\bf 84}, 113009 (2011)
  [arXiv:1109.6183 [hep-ph]].

\bibitem{Aaltonen:2008ah} 
  T.~Aaltonen {\it et al.}  [CDF Collaboration],
  Phys.\ Rev.\ Lett.\  {\bf 102}, 091805 (2009)
  [arXiv:0811.0053 [hep-ex]].

\bibitem{Dvergsnes:2004tw} 
  E.~W.~Dvergsnes, P.~Osland, A.~A.~Pankov and N.~Paver,
  Phys.\ Rev.\ D {\bf 69}, 115001 (2004)
  [hep-ph/0401199].

\bibitem{nicrosini} For reviews see, e.g.,
O.~Nicrosini and L.~Trentadue, in {\it Radiative Corrections for
$e^+e^-$ Collisions}, ed. J.~H.~K\"uhn 25 (Springer, Berlin,
1989), p. 25; in {\it QED Structure Functions, Ann Arbor, MI,
1989}, ed. G.~Bonvicini, AIP Conf. Proc. No. 201 (AIP, New York,
1990), p. 12.

\bibitem{physicsatlep2}
For a review see, e.g., W.~Beenakker and F.~A.~Berends: {\it Proc.
of the Workshop on Physics at LEP2}, CERN 96-01, vol. 1, p. 79 and
references therein.

\bibitem{Pankov:2005kd}
 A.~A.~Pankov, N.~Paver and A.~V.~Tsytrinov,
 Phys.\ Rev.\ D {\bf 73}, 115005 (2006)
 [hep-ph/0512131].

\end{thebibliography}
\end{document}